\documentclass[10pt,submission]{dmtcs}
\usepackage[a4paper]{geometry}

\usepackage{amsmath,amsfonts,amssymb}
\usepackage{xcolor}
\usepackage{epsf}
\usepackage{latexsym}
\usepackage{comment}
\usepackage{psfrag}

\usepackage{amssymb}
\usepackage{epsfig}
\usepackage{pstricks}

\usepackage{fancybox}
\usepackage{gastex}

\usepackage[latin1]{inputenc}
\usepackage{comment}

\newtheorem{theorem}{Theorem}
\newtheorem{lemma}{Lemma}
\newtheorem{example}{Example}
\newtheorem{definition}{Definition}

{\begin{center}\begin{Sbox}\begin{minipage}{\linewidth}\begin{small}\sf\color{red}(#1)}%
{\end{small}\end{minipage}\end{Sbox}\end{center}\colorbox{yellow!90!black}{\TheSbox}}
{\begin{center}\begin{Sbox}\begin{minipage}{\linewidth}\begin{small}\sf\color{black}(#1)}%
{\end{small}\end{minipage}\end{Sbox}\end{center}\colorbox{green!90!black}{\TheSbox}}

\DeclareMathOperator{\Occ}{Occ}

\DeclareMathOperator{\Pref}{Pref}
\newcommand\len[1]{{\left\lvert#1\right\rvert}}

\newcommand{\alphabet}{\mathcal{A}}
\newcommand{\auto}{\mathcal{T}}
\def\Ex{\mathbf{E}}

\def\bK{\mathbb{K}}
\def\bG{\mathbb{G}}

\def\bM{\mathbb{M}}
\def\bS{\mathbb{S}}

\def\Pr{\mathbf{P}}
\def\alph{{\mathcal{A}}}
\def\MR{{\mathcal{R}}}
\def\MU{{\mathcal{U}}}
\def\MM{{\mathcal{M}}}
\def\MD{{\mathcal{D}}}
\def\MN{{\mathcal{N}}}

\def\ME{{\mathcal{E}}}

\def\MA{{\mathcal{A}}}

\def\KK{{\mathfrak{K}}}

\def\MO1{{\mathcal{O}^1}}

\def\MM{{\mathcal{M}}}

\def\MN{{\mathcal{N}}}

\def\Ib{{\mathbf{I}}}
\def\1b{{\mathbf{1}}}

\def\b2{{\overline{2}}}

\def\bm1k{{\overline{k\!-\!1}}}

\def\MN{{\mathcal{N}}}
\def\MU{{\mathcal{U}}}
\def\MC{{\mathcal{C}}}
\def\MCc{{\mathcal{C}_{\circ}}}

\def\MK{{\mathcal{K}}}
\def\MA{{\mathcal{A}}}
\def\ML{{\mathcal{L}}}
\def\MAstar{\MA^{\star}}

\title{Constructions for Clumps Statistics}
\author{F. Bassino\addressmark{1}, J. Cl\'ement\addressmark{2}, J.
  Fayolle\addressmark{3}, \and P. Nicod\`eme\addressmark{4}}
\address{\addressmark{1}IGM, Universit\'e de Marne la Vall\'ee, 77454 Marne-la-Vall\'ee Cedex 2, 
France. \email{\footnotesize \tt Frederique.Bassino@univ-mlv.fr}\\
  \addressmark{2}GREYC, CNRS-UMR 6072, Universit\'e de Caen, 14032 Caen, France. 
\email{\footnotesize \tt Julien.Clement@info.unicaen.fr}\\
  \addressmark{3}LRI; Univ. Paris-Sud, CNRS ; B\^at 490, 91405 Orsay, France. 
\email{\footnotesize\tt Julien.Fayolle@lri.fr}\\
  \addressmark{4}LIX, CNRS-UMR 7161, \'Ecole polytechnique, 91128,
  Palaiseau, France. \email{\footnotesize\tt
    nicodeme@lix.polytechnique.fr}}

\begin{document}

\maketitle

\begin{abstract}
We consider a component of the word statistics known as clump;
starting from a finite set of words, clumps are maximal overlapping
sets of these occurrences. This parameter has first been studied by
Schbath~\cite{Schbath1995a} with the aim of counting the number of occurrences
of words in random texts. Later work with  similar probabilistic approach
used the Chen-Stein approximation for a compound Poisson distribution, where
the number of clumps follows a law close to Poisson.
Presently there is no combinatorial counterpart to this approach, and we 
fill the gap here. We emphasize the fact that, in contrast with
the probabilistic approach which only provides asymptotic results,
the combinatorial approach provides exact results that are useful when considering short sequences.
\end{abstract}

\section{Introduction}
Counting words and motifs in random texts has provided extended studies
with theoretical and practical reasons. 
Much of the present combinatorial research has built over the work of Guibas and 
Odlyzko~\cite{GuiOdl81a,GuiOdl81b}
who defined the autocorrelation polynomial of a word. As an apparently surprising consequence
of their work, the waiting time for the first occurrence of the word $111$ in a Bernoulli
string with probability $1/2$ for zeroes and ones is larger than the waiting time
for the first occurrence of the word $100$. This is due to the fact that the words $111$ occur
by {\sl clumps} of ones, the probability of extending a clump by one position being $1/2$;
this implies that the average number of $111$ in a clump is larger than one; in contrast,
there is only one $100$ in each clump of $100$. Since the probability that the word $111$
and the word $100$ start at a given position both are $1/8$, the interarrival time
of clumps of $111$ is larger than the interarrival time of clumps of $100$.

We analyze in this article several statictics connected to clumps of one word
or of a reduced set of words.
Our approach is based on properties of the R{\'e}gnier-Szpankowski~\cite{ReSz98b}
decomposition of languages along occurrences of the considered
word or set of words and on properties of the prefix codes generating the clumps.
We provide explicit generating functions in the Bernoulli model
for statistics such as (i) the number of clumps, (ii) the number of
$k$-clumps,
(iii) the number of positions of the texts covered by clumps, and (iv)
the size of clumps in infinite texts;
these results may be extended to a Markov model, providing some
technicalities.
We consider also in the Bernoulli model an algorithmic approach where we construct
deterministic finite automatas recognizing clumps.
This approach extends directly to the Markov model, and we obtain
as a direct consequence a Gaussian limit law for the number of clumps
in random texts.

Consider a rough first approximation for clumps of one word.
If the probability occurrence of a word $w$
is small, the probability of clumps $\KK$ of this word is small. This implies that
the number of clumps in texts of size $n$ follows a Poisson law of parameter
$\lambda=n\times \Pr(\mbox{a clump starts at position}\ i)$, where $i$
is a random position. Approximating further, the random number of occurrences $\Omega$
of the word $w$ in a clump follows a geometric law with parameter $\omega$, where $\omega$
is the probability of self-overlap of the word.
Schbath and Reinert~\cite{ReiSch1998} obtained in the Markov case
of any order a coumpound Poisson limit law for the
 count of number of occurrences by the Chein-Stein method.
See Reinert {\it et al.}~\cite{ReiSchWat2000} for a review
and Barbour {\it et al.}~\cite{BarHolJan1992b}
for an extensive introduction to the Poisson approximation.
Schbath~\cite{Schbath1995a} give the first moment of the number of $k$-clumps and
of the number of clumps in Bernoulli texts. Recently, Stefanov {\it et al.}~\cite{SteRobSch07}
use a stopping time method to compute the distribution of clumps; their results are
not explicit and practical application
of their method requires the inversion of a probability generating function.

We describe in Section~\ref{sec:prelim} our notations and the R\'egnier-Szpankowski
language decomposition.
Section~\ref{sec:cluone} and Section~\ref{sec:clumul} respectively
provide our analysis in the case of counting clumps and $k$-clumps of one word and of
a finite set of words.
We prove by an automaton construction a normal limit law for the number of clumps in Section~\ref{sec:limlaw}

\section{Preliminaries}
\label{sec:prelim}
We consider a finite alphabet $\mathcal{A}$.
Unless explicitely stated when considering a Markov source,
 the texts are generated by a non-uniform Bernoulli source over the alphabet $\mathcal{A}$.
Given a set of words, clumps of these words may be seen as a generalization  of runs of one letter.
\paragraph{Clumps and $k$-clumps.} 
When considering  a reduced set of words $U=\{u_1,\dots,u_r\}$ where each word $u_i$ 
has size at least $2$,
 a clump is a maximal set of occurrences
of words of $U$ such that
\begin{itemize}
\item any two consecutive letters of the clump belong to (is a factor of)
at least one occurrence,
\item either the clump is composed of a single occurrence that overlaps no other occurrences,
or each occurrence {\sl overlaps} at least one other occurrence.
\end{itemize}
This definition naturally applies also to the case where $U$ is composed of a single word.

As example,
considering the set $U=\{aba,bba\}$ and the text $T=bbbabababababbbbabaababb$,
we have
\[ T=b\underline{bbababababa}bb\underline{bbaba}\,\underline{aba}bb\]
where the clumps are underlined.
The word $bbababababa$ beginning at the second position of the text is a clump, 
and so are the words $bbaba$ and $aba$ beginning at the 15th and 20th positions.
On the contrary, the word $ababa$ beginning at the sixth position is not a clump 
since it is not maximal; neither is a clump the word $bbabaaba$ beginning at the 15th position,
 since its two-letters factor $aa$ is neither a factor of an occurrence of $aba$
nor of an occurrence of $bba$.

More formally, we use as an intermediate step {\sl clusters}, following 
Goulden and Jackson~\cite{GouldenJackson1983}.
\begin{definition}[Clumps]
A {\rm clustering-word} for the set $\mathcal{U}=\{u_1, \dots, u_r\}$ is a word $w \in
\alph^*$ such that any two consecutive positions in $w$ are {\rm covered} by
the same occurrence in $w$ of a word $u \in \mathcal{U}$. The position $i$ of the word $w$ is {\rm covered} by a word $u$ if $u=w[(j-|u|+1)\dots j]$ for some $j \in \{|u|,\dots,n\}$ and $j-|u|+1\leq i\leq j$. 
A {\rm cluster} of a clustering-word $w$ in $\mathcal{K}_\mathcal{U}$ is a set of
occurrence positions subsets
$
\{\, \mathcal{S}_u \subset \Occ(u,w)\, | \ u\in \mathcal{U} \,\}
$
which covers
exactly $w$, that is,
every two consecutive positions $i$ and $i+1$ in $w$ are
covered by at least one same occurrence of some $u\in
\mathcal{U}$. More formally
\[
\forall i \in \{1,\dots,\len{w}\!-\!1\}\quad \exists\, u \in \mathcal{U},\, \exists\, p \in
\mathcal{S}_u\quad\mbox{{\rm such that}}\quad  p-\len{u}+1<i+1 \le p.
\]
A {\rm clump}, generically denoted here by $\KK$ is a maximal cluster in the sense that there exists no
occurrence of the set $\mathcal{U}$ that overlaps the corresponding
clustering word without being a factor of it.
\end{definition}
Note that a single word is a cluster and that, as mentionned previously,
 a clump may be composed of a single word.

A {\sl $k$-clump} of occurrences of $U$
(denoted by $\KK^{(k)}$) is a clump containing exactly $k$ occurrences of $U$.

We aim here at providing explicit analytic formulas 
for the moments of the number of clumps, the total
size of text covered by clumps or the number of clumps with exactly $k$
occurrences.

\paragraph{Notations.}
We consider the {\sl residual} language $\MD=\ML.w^{-}$ as $\MD=\{x, \ x.w\in \ML\}$.

In case of ambiguity, we will use a 
bracket notation $\{\ML\}(z,\dots)$ to represent the generating function 
of the language $\{\ML\}$; in particular, for $\MD=\ML.w^{-}$, 
we write $\{\ML.w^{-}\}(z,\dots)=\MD(z,\dots)$.

Considering two languages $\ML_1$ and $\ML_2$, if we have
$\ML_1\subset\ML_2$,
we write $\ML_2-\ML_1=\ML_2\setminus \ML_1$ as the difference of sets;

\paragraph{Reduced set of words.} A set of words $U=\{u_1,\dots,u_r\}$ is reduced if no $u_i$
is factor of a $u_j$ with $i$ different of $j$.
\paragraph{Autocorrelations, correlations and right extension sets of words.}
We recall here the definition of {\sl Right Extension Set} introduced
in Bassino {\it et al.}~\cite{BaClFaNi07}.

The {\rm right extension set} of a pair of words ($h_1, h_2$) is
\[
\mathcal{E}_{h_1, h_2} = \{\  e \quad | \quad
\mbox{there exists} \  e' \in \alph^+ \quad \mbox{such that} \quad 
h_1 e =e'h_2 \quad \mbox{with} \ 0 < \len{e} < \len{h_2} \}.
\]  
If the word $h_1$ is not factor of $h_2$ this extension set of
$h_1$ to $h_2$ is the usual correlation set of $h_1$ and $h_2$

When we have $h_1=h_2$, we get the autocorrelation set $\mathcal{C}_{h,h}$ of the word $h$
that we will note further $\mathcal{C}$ when there is no ambiguity.

We also note $\MCc=\MC-\epsilon$. Remark that $\MCc$ is empty
if the word $w$ has no autocorrelation.

We remark here that the empty word $\epsilon$ belongs to the autocorrelation set of a word.
Note also that the correlation set of two words may be empty.

We have as examples
\[ \mathcal{C}_{aabaa,aab}=\{b,ab\},\qquad  \mathcal{C}_{ababa,ababa}=\{\epsilon,ba,baba\}.\]

\paragraph{Generating functions.}
We aim at computing the number of a given object in random texts by use
of generating functions such as
\begin{equation}
\label{eq:gf}
L_v(z,x)=\sum_{T\in \ML}\Pr(T)z^{|T|}x^{|T|_v}= \sum l_{n,i}x^i z^n
\end{equation}
where $|T|_v$ is the number of occurrences of the object $v$ in the text $T$
and $l_{n,i}$ is the probability that a text of size $n$ has $i$ occurrences
of this object. This extends naturally for counting more than one object
by considering multivariate generating functions with several parameters.

If the random variable $X_n$ counts the number of objects in a text of size $n$,
we get from Equation (\ref{eq:gf}) 
\[\Ex(X_n)=[z^n]\left.\frac{\partial L(z,x)}{\partial x}\right|_{x=1},\qquad
\Ex(X_n^2)=[z^n]\left.\frac{\partial }{\partial x}x\frac{\partial L(z,x)}{\partial x}\right|_{x=1}.\]
Recovering exactly or asymptotically these moments follows then from classical
methods.

\section{Formal language approach}
\subsection{R\'egnier and Szpankowski decomposition}
Since our work extends the formal language approach of 
R\'egnier and Szpankowski~\cite{ReSz98b}, 
we recall it  here.

Considering one word $w$, R\'egnier and Szpankowski 
use a natural parsing or decomposition of texts with at least one occurrence of $w$, where
\begin{itemize}
\item there is a first occurrence at the right extremity of a ``subtext'', the set of which
constitute a {\sl Right} language,
\item possibly followed by other occurrences, that are separated by ``subtexts'' that
constitute the {\sl Minimal} language,
\item and completed by ``subtexts'' that provide no other occurrences.
\end{itemize}
Moreover, there is a language without any match of the considered word $w$.
R\'egnier~\cite{Regnier00b}, further extended this approach to a reduced set of words.

We follow here the book of Lothaire~\cite{Lot05}(Chapter 7) which presents their method.

We consider a set of words $V=\{v_1,\dots,v_r\}$. We have, formally
\begin{definition}{Right, Minimal, Ultimate and Not languages}
\label{def:RMUlang}
\begin{itemize}
\item The :''Right'' language $\mathcal{R}_i$ associated to the word $v_i$ is the set of words
\[ \mathcal{R}_i=\{r\ |\quad r=e.v_i\quad\mbox{and}\quad \not\exists e'\in V,\  r=xe'y, \ |y|>0\}.\]
\item The ``Minimal'' language $\mathcal{M}_{ij}$ leading from a word $v_i$ to a word $v_j$ is the set of words
\[ \mathcal{M}_{ij}=\{m\ |\quad v_i.m=e.v_j\quad\mbox{and}\not\exists e'\in V,\  v_i.m=xe'y, \ |x|>0,|y|>0\}.\]
\item The ``Ultimate'' language completing a text after an occurrence of the word $v_i$ is the set of words
\[ \mathcal{U}_{i}=\{u\ |\quad \not\exists e\in V,\  v_i.u=xey, \ |x|>0\}.\]
\item The ``Not'' language completing a text after an occurrence of the word $v_i$ is the set of words
\[ \mathcal{N}=\{n\ |\quad \not\exists e\in V,\  n=xey\}.\]
\end{itemize}
\end{definition}
The notations $\MR, \MM$, $\MU$ and $\MN$ refer here to
the Right, Minimal, Ultimate and Not languages
of a single word.

Considering as example the word $w=ababa$; in the following texts, the
underlined words belong to the set $\MM$; the overlined text does not
since the word represented in bold faces is an intermediate occurrence.
\[ababa\underline{aaaaababa}\qquad ab{\mathbf{aba}}\overline{{\mathbf
{ba}}bbbbbbbb}\qquad ababa\underline{ba}.\]
Considering the matrix $\bM$ such that $\bM_{ij}=\MM_{ij}$, we have
\begin{eqnarray}
\label{eq:lotvw1}
&\bigcup_{k\geq 1}\left(\bM^k\right)_{i,j}=\MA^{\star}\cdot w_j +
\MC_{ij}-\delta_{ij}\epsilon,
\quad &\MU_i\cdot\MA=\bigcup_j \MM_{ij}+\MU_i-\epsilon, \\
\label{eq:lotvw5}
&\MA\cdot \MR_j -\left(\MR_j-w_j\right)=\bigcup_i w_i\MM_{ij},
\quad &\MN\cdot w_j= \MR_j+\bigcup_i
\MR_i\left(\MC_{ij}-\delta_{ij}\epsilon\right).
\end{eqnarray}
If the size of the texts is counted by the variable $z$ and
the occurrences of the words $v_1,\dots,v_r$ are counted respectively by $x_1,\dots,x_r$, we get the
matrix equation
\begin{equation}
\label{eq:matnofac}
F(z,x_1,\dots,x_r)= \MN(z)+(x_1\MR_1(z),\dots,x_r\MR_r(z))\big(\Ib-\bM(z,x_1,\dots,x_r)\big)^{-1}
\left(\begin{array}{c}\MU_1(z)\\\vdots\\\MU_r(z)\end{array}\right).
\end{equation}
In this last equation, we have $\bM_{ij}(z,x_1,\dots,x_r)=x_j\MM_{ij}(z)$
and the generating functions $\MR_i(z)$, $\MM_{ij}(z)$, $U_j(z)$ and $\MN(z)$ can be
computed explicitly from the set of Equations (\ref{eq:lotvw1},~\ref{eq:lotvw5}).

In particular, when considering the Bernoulli weighted case $A(z)=z$ 
and a single word $w$ with $\pi_w=\Pr(w)$, 
we have the set of equations
\small
\begin{equation}
\label{eq:RMUN}
R(z)=\dfrac{\pi_w z^{|w|}}{D(z)},\ 
M(z)=1+\dfrac{z-1}{D(z)},\ 
U(z)=\dfrac{1}{D(z)},\ 
N(z)=\dfrac{C(z)}{D(z)}\quad
\left(\dfrac{1}{D(z)}=\dfrac{1}{\pi_w z^{|w|} +(1-z)C(z)}\right)
\end{equation}
\normalsize
\begin{equation}
\label{eq:LFzx}
\MA^{\star}=\MN+\MR \MM^{\star}\MU \quad \Longrightarrow \quad F(z,x)=\dfrac{1}{1-z+\pi_w z^{|w|}\dfrac{1-x}{x+(1-x)C(z)}}=\sum_{n,k}f_{n,k}x^kz^n.
\end{equation}
In this last equation, $f_{n,k}$ is the probability that a text of size $n$ has $k$ occurrences of $w$.

\subsection{Clump analysis for one word}
\label{sec:cluone}
The decomposition of R\'egnier and Szpankowski is based on a parsing by the occurrences
of the considered words. We use a similar approach, but parse with respect to the occurrences
of clumps. As a major difference, when they consider the minimal language separating
two occurrences, these two occurrences may overlap; in contrast, by definition, overlapping
of clumps is forbidden.

A clump of the word $w$ is basically defined as $w\MC^{\star}$, since any element of $\MCc$
concatanated to a cluster extends this cluster.

Considering the word $w=aaa$, we have $\MC=\{\epsilon,a,aa\}$ and $\MC^{\star}$
is ambiguous.
We can however generate unambiguously $\MC^{\star}$ as described in the next section.

\subsubsection{A prefix code $\MK$ to generate unambiguously $\MC^{\star}$}
\label{subsec:unambpref}
Since $\MCc$ is a finite language, it is possible to find a
prefix code $\MK$ generating $\MCc$; moreover, for $c_1,c_2
\in \MC-\epsilon$ and $|c_1|<|c_2|$, the word $c_1$ is
a proper suffix of $c_2$. Otherwise stated, the prefix code
$\MK=\{\kappa_1,\dots,\kappa_k\}$ is built over words $q_1,q_2,\dots,q_k$ and may be written as
$\MK=\{q_1,q_2 q_1, \dots, q_k q_{k-1}\dots q_1\}$.

We Refer to Berstel and Perrin~\cite{BerPer85} for an introduction to prefix codes.
See also Berstel~\cite{Ber05} for an analysis of counts of words of the pattern $U$
by semaphore codes $U\,-\MA^{\star}U\MA^+$.
We have the following lemma
\begin{lemma}
\label{lem:MK}
The prefix code $\MK=\MCc \,\backslash\, \MCc \MA^+$ generates unambigously the 
language $\MC^{\star}$.
\end{lemma}
\begin{proof}
It is clear that $\MK$ is prefix. Consider $w\in \MCc-\MK$ if this last set is not empty.
Since $\MK$ is a set of words of $\MC$ without any prefix in $\MC$, we have {\it a contrario}
$w=u.v$ with $u$ and $v$ non-empty and in $\MC$. We have $|u|<|w|$
and $|v|<|w|$; if $u$ or $v$ does not belong to $\MK$, we may iterate the process 
on the corresponding word. Since $|w|$ is finite, after a finite number
of steps, we get to a decomposition $w=\kappa_{i_1}\dots\kappa_{i_j}$
where each $\kappa_i$ is in $\MK$. Since $\MK$ is a code, the decomposition of each word
of $\MC$ over $\MK$ is unique and so is the decomposition of any word of $\MC^{\star}$.
\end{proof}
\begin{example}
Let $w=abaabaaba$. We have
\[
\begin{array}{l}
abaabaaba|\epsilon\\
\phantom{aba}abaaba|aba\\
\phantom{abaaba}aba|abaaba\\
\phantom{abaabaab}a|baabaaba\\
\end{array} \quad \Longrightarrow \quad
\MC=\{\epsilon,aba,abaaba,baabaaba\} \quad \Longrightarrow \quad\MK=\{aba,baabaaba\}.
\]
\end{example}
The {\sl periods} of a word $w$ is the set of integers $\{|h|,h\in\MCc\}$; the irreducible
periods is the subset of periods of which all the periods may be deduced.
As follows from Guibas and Odlyzko~\cite{GuiOdl81a} and Rivals and Rahmann~\cite{RivRah03},
when considering the word $ababaccababa$, the irreducible periods are $7,9$ while the period $11$ can be deduced from the periods $7$ and $9$. However, we have here $\MK=\MC=\{ccababa,baccababa,babaccababa\}$, which implies somehow against intuition that, in general, 
 there is no bijection between the irreducible periods and the prefix code of a word.

\paragraph{Constructing the prefix-code $\MK$.}
We use the following algorithm:
\begin{enumerate}
\item start with the word $w$;
\item shift $w$ to the right to the first self-overlapping
position;
 Let $\kappa_1$ be the trailing suffix so obtained; insert it in a
trie $\ME$;
\item repeat shifting, obtaining new trailing suffixes; for each new
suffix generated, try an insertion in the trie. If you reach a leaf,
drop the suffix; elsewhere insert it.
\end{enumerate}
The worst case complexity for this construction is $O(|w|)$, but the average complexity is
$O(|\MK|\log(|\MK|))$, the average path length of a trie built over $|\MK|$ keys.

\subsubsection{The language decomposition}
Considering the word $w=aaaaa$, we have $\MC=\{a,aa,aaa,aaaa\}$ and
$\MK=\{a\}$. Moreover, we have
$\MM=\{a,b(b+ab+aab+aaab+aaaab)^{\star}aaaaa\}$. We get here
$\MK\subset \MM$ and $\MM - \MK = \ML w$;
The language $\MM$  and $\MK$ are indeed connected by a simple property that we describe now.
\begin{lemma}
\label{lemma:MminusK}
For any word $w$ with autocorrelation set $\MC$, prefix code $\MK$
generating $\MC^{\star}$ and minimal language $\MM$, there exists
a non-empty language $\ML$ such that
\begin{equation}
\label{eq:MminusK}
  \MK\subset\MM\qquad \mbox{and}\qquad\MM-\MK=\ML w.
\end{equation}
\end{lemma}
\begin{proof}
We have $\MK\subset\MC$ and $\MK\subset\MM$; therefore, we have
$\MK\subset \MM\cap\MC$. We prove that if $w\in \MC - \MK$ then $w \not\in \MM$. Let us suppose that $w \neq \epsilon$ and $w\in \MC -\MK$.
This implies that $w\in \MK\MA^{\star}$ by definition of $\MK$.
Therefore, we have $w=\kappa v$ with $\kappa\in \MK$ and $|v|>0$.
As a consequence, $w$ cannot belong to the minimal language $\MM$,
the word $\kappa$ corresponding to a previous occurrence of $w$.
\end{proof}

This leads immediately to the fundamental lemma.
\begin{lemma}
The basic equation for the combinatorial decomposition of texts on the alphabet $\mathcal{A}$ where $v$ counts some object in the clump of a word $w$ is
\begin{equation}
\label{eq:basic}
\MAstar_v=\MN + \MR w^-(w\MC^{\star})_v\big((\MM-\MK)w^-(w\MC^{\star})_v\big)^{\star}\MU,
\end{equation}
\end{lemma}
\begin{proof}
The Equation (\ref{eq:basic}) follows from the parsing
\begin{itemize}
\setlength{\itemsep}{-2pt}
\item either there is no occurrence of $w$, the Not language $\MN$,
\item or
   \begin{enumerate}
    \setlength{\itemsep}{-2pt}
    \item we read until the first occurrence : $\MR w^- w$,
    \item followed by any number of overlapping occurrences of $w$ (a clump less the first occurence): $\MC^{\star}$,
    \item followed by any number of
       \begin{enumerate}
         \setlength{\itemsep}{-0pt}
         \item next occurrence of $w$ without overlap: $(\MM-\MK)w^- w$
         \item and any number of overlapping occurrences of $w$: $\MC^{\star}$.
       \end{enumerate}
   \end{enumerate}
\end{itemize}
\end{proof}

We can now use the preceeding lemma to count several parameters related to the clumps.

\subsubsection{Counting parameters related to the clumps}
\label{subsec:count}
Let $\KK(z,x,t)$ be the generating function where the variable $x$ counts the number
of occurrence of $w$ in a clump, and the variable $t$ counts
the size of the clumps; the variable $z$
is used here to count the total length of the texts. We also use a variable $u$ to count the number
of clumps. We have the following theorem
\begin{theorem}
In the weighted model such that $A(z)=z$, the generating function counting the number of occurrences of a word $w$ and the number of positions covered by the clumps of $w$ verifies
\begin{equation}
\label{eq:gfbasic}
F(z,\KK(z,x,t))=\MN(z)+\frac{\MR(z)}{\pi_w z^{|w|}}\KK(zt,x)\frac{1}{1-\dfrac{\MM(z)-\MK(z)}{\pi_w z^{|w|}}\times \KK(zt,x)}\,\MU(z)
\end{equation}
where the generating function of the clumps verifies
\begin{equation}
\label{eq:Kappa}
\KK(z,x,t)=x \pi_w(zt)^{|w|}\frac{1}{1-x \MK(zt)}
\end{equation}
As a consequence, the generating function counting also the number of clumps is
\begin{equation}
\label{eq:allvar}
G(z,x,t,u)=F(z,u\KK(z,x,t)).
\end{equation}
\end{theorem}
\begin{proof}
This theorem follows from Lemma (\ref{lem:MK}) and from a direct translation of Equation (\ref{eq:basic}) into
generating functions.
\end{proof}

\subsubsection{Occurrences of clumps.}
Considering $G(z,u\KK(z,1,1))$ 
in Equation (\ref{eq:gfbasic}) and using Equation (\ref{eq:Kappa})
provides the generating function
\begin{equation}
\label{eq:gfU}
O^{(\gamma)}(z,u)  =\sum_{n,i}o^{(\gamma)}_{n,i}u^i z^n 
= \MN(z)+
\frac{u\MR(z)\MU(z)}{1-u\MM(z)+(u-1)\MK(z)}
\end{equation}
where $o^{(\gamma)}_{n,i}$ is the probability of getting $i$ clumps (of any size)
in a text of size $n$. Considering $\Gamma_n$, the expectation of
number of clumps in texts of size $n$, we get by differentiation
\[\sum_n \Gamma_n z^n =
\frac{\MR(z)\MU(z)(1-\MK(z))}{(1-\MM(z))^2}=\frac{\pi_w
z^{|w|}(1-\MK(z))}{(1-z)^2}.\] 
This implies that $\Gamma_n=(n-|w|+1)\pi_w(1-\MK(1))-\pi_w\MK'(1)$, to
compare with the expectation $(n-|w|+1)\pi_w$ of the numerb of
occurrences of the word $w$.

\subsubsection{Occurrences of $k$-clumps.}
By considering the equation of a clump of occurrences of $w$, we can write
\[w\MC^{\star}=w+w\MK+w\MK^2+\dots (v-1+1)w\MK^{k-1}+\dots\]
to count clumps with exactly $k$ occurrences of $w$.

Writing $\KK^{(k)}(z,v)$ the generating function which counts with the variable
 $z$ the size of the clumps
and where
the variable $v$ selects $k$-clumps, we have
\[\KK^{(k)}(z,v)=\pi_w z^{|w|}\left(\frac{1}{1-\MK(z)}+(v-1)\MK(z)^{k-1}\right)\]
Substituting this in Equation~\ref{eq:gfbasic} 
gives
\[O^{(\gamma_k)}(z,v)= \sum o^{(\gamma_k)}_{n,i} v^i z^n =
\MN(z)+\frac{\MR(z)}{\pi_w z^{|w|}}\KK^{(k)}(z,v)\frac{1}{1-\dfrac{\MM(z)-\MK(z)}{\pi_w z^{|w|}}\times
\KK^{(k)}(z,v)}\,\MU(z),
\]
where $o^{(\gamma_k)}_{n,i}$ is the probability that a text of size $n$
contains exactly $i$ $k$-clumps.

\subsubsection{Probability that a random position is covered by a clump}
This follows from the knowledge of the number of positions of the texts covered
by the clumps.

Let $P_n$ be the random variable counting
the number of positions covered by the clumps of a word $w$ in texts of size $n$
and $H_n$ be the probability that a random position is covered by a clump in a text of size $n$.

Let $F(z,t)=G(z,\KK(zt,1))$ where $G(z,\KK)$ is given by Equation (\ref{eq:gfbasic})
be the generating function  counting the size of the texts and
the number of positions covered by clumps. 
We have
\[ H_n= \sum_{i\geq 0} \frac{i}{n}\Pr(P_n=i)\qquad\Longleftrightarrow\qquad 
      H_n=[z^n]\left.\frac{\partial}{\partial t}z\int_0^z F(y,t)dy\right|_{t=1}.\]

\subsection{Clumps of a finite set of words}
\label{sec:clumul}
We provide in this section a matricial solution for counting clumps of a reduced finite
set of words.
For simplicity sake we consider a set of two words $w_1$ and $w_2$ but our approach is amenable to any reduced finite set.

Similarly to the one word case, we are lead to consider prefix codes generating the
correlation of two words. Writing $\MC_{ij}^{\star}$ with $i\neq j$ makes no sense
in terms of language decomposition. However, we can write as previously $\MK_{ij}=\MC_{ij}-\MC_{ij}\MA^{+}$, which defines a minimal correlation language with good properties.

We have as examples
\begin{example}
Let $w_1=aabaa$ and $w2=aaa$. We have $\MC_{12}=\{a,aa\}$ and
$\MK_{12}=\{a\}$. In this case, we have $\MC_{12}=\MC_{22}-\{\epsilon\}$
and $\MK_{12}=\MK_{22}$.
\end{example}
\begin{example}
Let $w_1=abab$ and $w2=baba$. We have $\MC_{12}=\MK_{12}=\{a,aba\}$. 
In this case, we have $\MC_{22}=\{\epsilon,ba\}$
and $\MK_{12}=a.\MK_{22}$.
\end{example}
Following a proof similar to the proof of Lemma (\ref{lemma:MminusK}),
there exists a language $\ML$ such that
\[\MM_{ij}-\MK_{ij}=\ML.w_j.\]
We can therefore write a minimal correlation matrix $\bK$, consider the matrix $\bS=\bK^{\star}$
and write a clump matrix $\bG$ as follows
\begin{equation}
\label{eq:clumul}
\bK=\left(\begin{array}{cc} 
               \MK_{11} & \MK_{12}\\ \MK_{21} & \MK_{22}\end{array}\right),\qquad
{\bS}=\bK^{\star}, \qquad  \bG=\left(\begin{array}{cc} 
               w_1\bS_{11} & w_1\bS_{12}\\ 
               w_2\bS_{21} & w_2\bS_{22}\end{array}\right)
\end{equation}
In this equation, $\bG_{ij}$ is a clump starting with the word $w_i$ and finishing with the word
$w_j$. We obtain now a fundamental matricial decomposition that can be used for further analysis,
\begin{align*}
\\[-6ex]
\MA^{\star}=(\MR_1 w_1^{-},\MR_2 w_2^{-})\,\bG\,\Big((\bM-\bK)^{-}\bG\Big)^{\!\!\!\begin{array}{l}\phantom{1}\\\star\end{array}}\!\!\left(\begin{array}{c}\MU_1\\\MU_2\end{array}\right)
\end{align*}
where we have $(\bM-\bK)^-_{ij}=(\MM_{ij}-\MK_{ij})w_j^-$.

\section{Automaton approach}
\label{sec:auto}
For a set $U=\{u_1, \dots, u_r\}$, we build a kind of ``Aho-Corasick''
automaton built on the following set of words $X$ 
\[
X= \{ u_i \cdot w \ | \ \text{$1\le i \le r$ and $w \in \{\epsilon\} \cup \mathcal{E}_{i, j}$ for some $j$}\}.
\]
The automaton $\auto$ is built on
$X$ with $Q=\Pref(X)$ (set of states), $i=\epsilon$ (initial
state). The transition function is defined (as in Aho-Corasick
construction) as 
\[
\delta(p, x) = \text{the longest suffix of $px \in
  \Pref(X)$}.
\] 
In order to count the number of clumps (for instance) the set of final states $T$ needs more attention: it is defined as 
\[
T = X \setminus X\alphabet^+.
\] 
This automaton accepts the language of
words ending by the first occurrence of a word in a clump.

We can easily derive from this automaton the generating function $f(z,
x_1, \dots , x_r, t, u)$ where $x_i$ marks an occurrence of $u_i$, $t$
marks the number of clumps, and $u$ the total length covered by the clump. 
Indeed, one has to mark some transitions in the adjacency
matrix $A$ according to some simple rules. 
\begin{itemize}
\item
To count occurrences of the
$u_i$'s, we have to mark with the formal variable $x_i$ transitions going to states
$\alphabet^*u_i \cap \Pref(X)$ \text{ (for $1 \le i \le r$)}. 
\item
For the number of clumps, on can mark
transitions going to states in $\mathcal{U}\setminus \mathcal{U} \alphabet^+=X \setminus X\alphabet^+$, that is
states corresponding to first occurences inside a clump.
\item Finally, for the total length covered by clumps. We have to put
  a formal weight on transitions going to a state $p \in
  \alphabet^*\mathcal{U} \cap \Pref(X)$ taking into account the number
  of symbols between the last occurrence of a word of $X$ and the new
  one at the end of $p$.  Let us define for a state $p$ (corresponding
  to a word with a occurrence of some word of $X$ at the end) the
  function $\ell(p)$ the maximal proper prefix $q$ of $p$ in
  $\alphabet^*\mathcal{U}$ if it exists or $\epsilon$ if there is no
  such prefix. Then we must mark all transitions going to $p$ with
  $u^{\len{p}-\len{\ell(p)}}$ (if $p \in \alphabet^*\mathcal{U} \cap
  \Pref(X)$). 
\end{itemize}

Of course the construction does not gives a minimal automaton. However the automaton is complete and deterministic so that the translation to generating function is straightforward.

\paragraph{Example}

\begin{enumerate}
\item
For one word $\mathcal{U}=\{u=bababa\}$, and $\mathcal{E}_u=\{ba, baba\}$. The set $X$ is 
\[
X=\{ bababa, babababa, bababababa\}.
\]

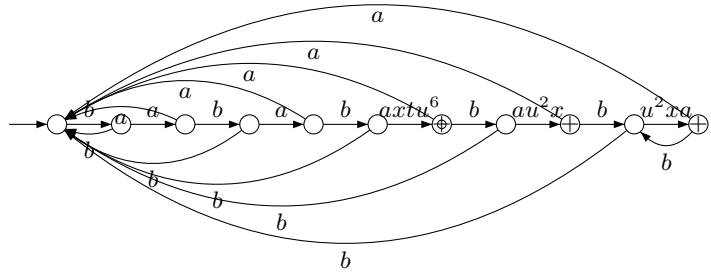
\begin{center}
\begin{small}
\setlength{\unitlength}{1.2pt}
\begin{picture}(222,88)(0,-88)
\gasset{Nw=6,Nh=6}
\node[Nmarks=i](n10)(12.24,-43.52){}
\node(n11)(32.24,-43.52){}
\node(n12)(52.24,-43.52){}
\node(n13)(72.24,-43.52){}
\node(n14)(92.24,-43.52){}
\node(n15)(112.24,-43.52){}
\node[NLangle=0.0,Nmarks=r](n16)(132.24,-43.52){$+$}
\node(n17)(152.24,-43.52){}
\node[NLangle=0.0](n18)(172.24,-43.52){$+$}
\node(n19)(192.24,-43.52){}
\node[NLangle=0.0](n20)(212.24,-43.52){$+$}
\drawedge(n10,n11){$b$}
\drawedge(n11,n12){$a$}
\drawedge(n12,n13){$b$}
\drawedge(n13,n14){$a$}
\drawedge(n14,n15){$b$}
\drawedge(n15,n16){$axtu^6$}
\drawedge(n16,n17){$b$}
\drawedge(n17,n18){$au^2x$}
\drawedge(n18,n19){$b$}
\drawedge(n19,n20){$u^2xa$}
\drawedge[curvedepth=3.07](n11,n10){$b$}
\drawedge[curvedepth=12.3](n13,n10){$b$}
\drawedge[curvedepth=18.85](n15,n10){$b$}
\drawedge[curvedepth=25.67](n17,n10){$b$}
\drawedge[curvedepth=37.43](n19,n10){$b$}
\drawedge[curvedepth=-5.61](n12,n10){$a$}
\drawedge[curvedepth=-13.77](n14,n10){$a$}
\drawedge[curvedepth=-19.18](n16,n10){$a$}
\drawedge[curvedepth=-26.13](n18,n10){$a$}
\drawedge[curvedepth=-37.62](n20,n10){$a$}
\drawedge[curvedepth=6.72](n20,n19){$b$}
\end{picture}
\end{small}
\end{center}
N.B.: The sign '$+$' on the automaton indicates that the corresponding prefix ends with some occurence of $\mathcal{U}$. The double oval states indicates the states where we know we have entered a new clump.

\item
For the set $\mathcal{U} = \{ u_1=aabaa, u_2=baab\}$ and the matrix of right extension sets is
\[
\mathcal{E} = \begin{pmatrix}
baa+abaa & b\\
aa & aab
\end{pmatrix}.
\]
The set $X$ is 
\[
X =\{
aabaa, aabaab, aabaabaa, aabaaabaa, baab, baabaa, baabaab\}.
\]
We have the following automaton (with $x$ and $y$  marking occurences respectively of $u_1$ and $u_2$. 
The automaton is complete and deterministic. However, for clarity's
sake, all transitions labelled by $a$ and $b$ ending respectively on
state $A$ and $B$ are omitted.  As before, the sign '$+$' indicates
that the corresponding prefix (or, equivalently, state) ends with some
occurence of $\mathcal{U}$. The double oval attribute indicates the
state where we know we have entered a new clump.
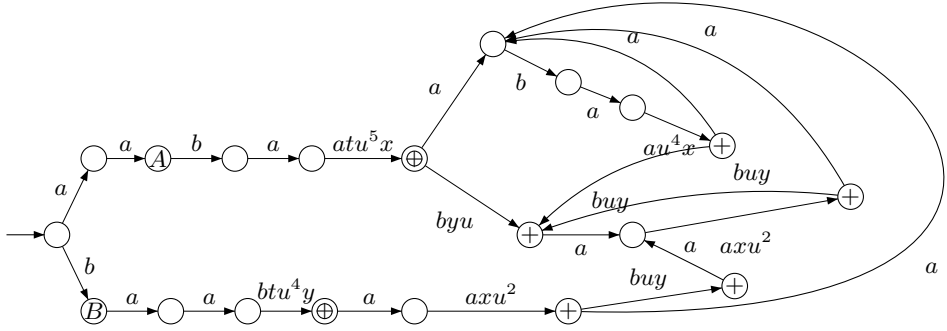
\begin{center}
\begin{small}
\setlength{\unitlength}{1.2pt}
\begin{picture}(300,120)(0,-120)
\gasset{Nh=8, Nw=8}
\node[Nmarks=i](n21)(8.75,-76.51){}

\node(n22)(20.26,-52.52){}

\node[NLangle=0.0](n23)(40.26,-52.52){$A$}

\node(n24)(64.26,-52.52){}

\node(n25)(88.26,-52.52){}

\node[NLangle=0.0,Nmarks=r](n26)(120.26,-52.52){$+$}

\node(n27)(144.75,-16.51){}

\node(n28)(168.26,-28.52){}

\node(n29)(188.26,-36.52){}

\node[NLangle=0.0](n30)(216.26,-48.52){$+$}

\node[NLangle=0.0](n31)(156.26,-76.52){$+$}

\node(n32)(188.26,-76.52){}

\node[NLangle=0.0](n33)(256.26,-64.52){$+$}

\drawedge[linewidth=0.28](n21,n22){$a$}

\drawedge[linewidth=0.28](n22,n23){$a$}

\drawedge[linewidth=0.28](n23,n24){$b$}

\drawedge[linewidth=0.28](n24,n25){$a$}

\drawedge[linewidth=0.28](n25,n26){$atu^5x$}

\drawedge[linewidth=0.28,ELdist=4.43](n26,n27){$a$}

\drawedge[linewidth=0.28,ELside=r,ELdist=3.58](n27,n28){$b$}

\drawedge[linewidth=0.28,ELside=r,ELdist=3.93](n28,n29){$a$}

\drawedge[linewidth=0.28,ELside=r,ELdist=0](n29,n30){$au^4x$}

\drawedge[linewidth=0.28,ELside=r,ELdist=3.78](n26,n31){$byu$}

\drawedge[linewidth=0.28,ELside=r,ELdist=3.21](n31,n32){$a$}

\drawedge[linewidth=0.28,ELside=r,ELdist=3.96](n32,n33){$axu^2$}

\node[NLangle=0.0](n34)(20.26,-100.52){$B$}

\node(n35)(44.26,-100.52){}

\node(n36)(68.26,-100.52){}

\node[NLangle=0.0,Nmarks=r](n37)(92.26,-100.52){$+$}

\node(n38)(120.26,-100.52){}

\node[NLangle=0.0](n39)(168.26,-100.52){$+$}

\node[NLangle=0.0](n40)(220.12,-92.61){$+$}

\drawedge[linewidth=0.28](n21,n34){$b$}

\drawedge[linewidth=0.28](n34,n35){$a$}

\drawedge[linewidth=0.28](n35,n36){$a$}

\drawedge[linewidth=0.28](n36,n37){$btu^4y$}

\drawedge[linewidth=0.28](n37,n38){$a$}

\drawedge[linewidth=0.28](n38,n39){$axu^2$}

\drawedge[linewidth=0.28,ELdist=2](n39,n40){$buy$}

\drawedge[ELside=r,ELdist=3.67,curvedepth=-13.71](n30,n27){$a$}

\drawedge[ELpos=60,curvedepth=-6.79,ELdist=2](n30,n31){$buy$}

\drawedge[ELside=r,ELpos=30,ELdist=2,curvedepth=-6.27](n33,n31){$buy$}

\drawedge[ELside=r,ELdist=4.59,curvedepth=-23.97](n33,n27){$a$}

\drawedge[ELside=r,ELdist=2.33,curvedepth=-0.25](n40,n32){$a$}

\drawbpedge[ELside=r,ELpos=27,ELdist=7.3](n39,-2,216.27,n27,27,124.72){$a$}

\end{picture}
\end{small}
\end{center}
\end{enumerate}

\section{Limit laws}
\label{sec:limlaw}
\subsection{Normal law}
A normal limit law for the number of clumps $U$ when
$U=O(n)$ in texts of size $n$ follows from the automaton
construction of Section~\ref{sec:auto}.
A Perron-Frobenius property asserts the existence
of a unique  dominant eigenvalue of the positive system; apply next a suitable
Cauchy integral
and large power Theorem of Hwang~\cite{Hwang94,Hwang96};
see~\cite{NiSaFl02}
for details.

\subsection{Poisson law for rare words}
In a Bernoulli model, if $\underline{p}$ and $\overline{p}$
are the minimal and maximal  probability of letters of the alphabet,
words of size $l<\frac{\log n}{\log(1/q)}$ have $O(n)$ number of
occurrences in texts of size $n$ with probability one. We consider
rare words with size over this threshold and number of occurrences
$O(1)$.
We prove in this case a Poisson-like limit law.
Taking a Taylor expansion of $O^{(\gamma)}(z,u)$ in Equation (\ref{eq:gfU}) at $u=0$, and
considering the $k$th Taylor coefficient, with $k=O(1)$
provide a rational generating function with respect to the
variable $z$ of the form 
\begin{equation}
\label{eq:Poisgf}
H_k(z)=[u^k]O^{(\gamma)}(z,u)=
\frac{\MR(z)\MU(z)(\MM(z)-\MK(z))^{k-1}}{(1-\MK(z))^k}
=\frac{\pi_w z^{|w|}\big(z-1+(1-\MK(z))D(z)\big)^{k-1}}{(1-\MK(z))^k(D(z))^{k+1}}.\end{equation}
We follow Fayolle~\cite{Fa04} to prove that the dominant root of the
denominator of this last equation is the smallest and positive root of
$D(z)=\pi_w z^{|w|}+(1-z)C(z)$; (see Equation (\ref{eq:RMUN})). 
Let $d$ be the smallest period of $w$. If $d\leq l/2$ classical results
about periods on words provide
$C(z)=1+\pi_uz^{|u|}+\dots +(\pi_uz^{|u|})^r+S(z)$ for a given word
$u$ with $|u|<l/2$, and $r\geq 2$; moreover $S(z)$ is a polynomial of
minimal degree at least $l/2$. Moreover, we have
$\MK(z)=\pi_uz^{|u|}+R(z)$
where $S(z)-R(z)$ is a polynomial with positive coefficients. This
entails that $|S(z)|$ and $|R(z)|$ are $o(1)$ for
$|z|<1/\overline{p}$.
Up to negligible terms, we get
\[|C(z)|=\left|\frac{1}{1-\pi_u z^{|u|}}\right|\geq
\frac{1}{1+\pi_u|z|^{|u|}}\geq \frac{1}{1+p|z|}\quad \mbox{for}\quad
|z|<\frac{1}{p}.\]
We also have $|1-\MK(z)|>0$ and $\pi_w z^{|w|}=o(1)$ for $|z|<1/p$. 
The Rouch\'e theorem in the disk $|z|<1/p$ the generating function 
$H_k(z)$ has a single pole which is a smallest modulus 
root $\rho$ of the equation $D(z)=0$. Perron-Frobenius considerations on the
automaton counting the number of occurrences of $w$ imply that this
pole is real positive.
A similar proof follows when $d> l/2$.

Writing $D(z)=Q(z)(1-z/\rho)$  and $P(z)=z-1+(1-\MK(z))D(z)$  we get as a first approximation
\[\Pr(O_n^{\gamma}=k)
     \approx \frac{\pi_w\rho^{|w|}}{Q(\rho)}\times
\frac{1}{k!}\left(\frac{\rho 
P(\rho)\times n}{(1-\MK(\rho))Q(\rho)}\right)^k\times \rho^{-n}.\]
A similar behaviour has been observed for occurrences of one word by
R\'egnier and Szpankowski~\cite{ReSz98b}.

\subsection{Length of the clumps in infinite texts}
Generating function of the size of the
clumps in infinite texts is  a sum of geometric random variables.

\section{Conclusion}
An interesting application of this article would be a combinatorial analysis
of {\sl tandem repeats} or multiple repeats that occur in genomes; large variations
of such repeats are characteristics of some genetic diseases.

Would it be possible to extend our approach to clumps of regular expressions?
We consider {\sl clumps of a regular expression} ({\it i.e.} contiguous
sets of positions such that each position is covered by at least one
word of the associated regular language and such that leading and
terminating positions of each occurrence is covered by at least two
occurrences). In this case the star-height theorem (CITE) inplies that
we cannot in general find a finite set of words $w_i$ and a finite set
of prefix codes $\MK_i$ with $1\leq i\leq \ell$ such that
the language $\bigcup_{1\leq i \leq n}w_i (\MK_i)^{\star}$
describes the clumps.

\nocite{Lot05,Regnier00b,ReSz98b,Hwang94,Hwang96,NiSaFl02,JacSzp94,FlaSeg95,ChoSchu63,GoJa79,GuiOdl81a,GuiOdl81b,Gantma59,Szpa01,SteRobSch07,ReiSchWat2000,Schbath1995a,BarHolJan1992b,RivRah03,Ber05,BerPer85}

\bibliographystyle{acm}
\bibliography{anaclump}

\end{document}